\documentclass[runningheads]{llncs}

\usepackage{graphicx}

\usepackage{multicol} 
\usepackage{paralist}
\usepackage{wrapfig} % for wrapping figures
\usepackage{color}
\usepackage{enumitem}

\newcommand{\quotes}[1]{``#1''}

\title{Modelling GDPR-Compliant Explanations for Trustworthy AI}

\author{Francesco Sovrano\inst{1} \and
	Fabio Vitali\inst{1} \and
	Monica Palmirani\inst{2}}
\authorrunning{F. Sovrano et al.}
\institute{DISI, University of Bologna \and
	CIRSFID, University of Bologna\\
	\email{\{francesco.sovrano2, fabio.vitali, monica.palmirani\}@unibo.it}}

\begin{document}
	
	\maketitle
		
	\begin{abstract}
		Through the General Data Protection Regulation (GDPR), the European Union has set out its vision for Automated Decision-Making (ADM) and AI, which must be reliable and human-centred. In particular we are interested on the Right to Explanation, that requires industry to produce explanations of ADM. The High-Level Expert Group on Artificial Intelligence (AI-HLEG), set up to support the implementation of this vision, has produced guidelines discussing the types of explanations that are appropriate for user-centred (interactive) Explanatory Tools. In this paper we propose our version of Explanatory Narratives (EN), based on user-centred concepts drawn from ISO 9241, as a model for user-centred explanations aligned with the GDPR and the AI-HLEG guidelines. Through the use of ENs we convert the problem of generating explanations for ADM into the identification of an appropriate path over an Explanatory Space, allowing explainees to interactively explore it and produce the explanation best suited to their needs. To this end we list suitable exploration heuristics, we study the properties and structure of explanations, and discuss the proposed model identifying its weaknesses and strengths.
	\end{abstract}
	
	\section{Introduction}
	The academic interest in Artificial Intelligence (AI) has grown together with the attention of countries and people toward the actual disruptive effects of Automated Decision Making (ADM \cite{eucommission16}) in industry and in the public administration, effects that may affect the lives of billions of people. Thus, GDPR (General Data Protection Regulation, UE 2016/679) stresses the importance of the Right to Explanation, with several expert groups, including those acting for the European Commission, have started asking the AI industry to adopt ethics code of conducts as quickly as possible \cite{cath2018artificial,floridi2018ai4people}.
	The GDPR draws a set of expectations to meet in order to guarantee the Right to Explanation (for more details see section \ref{sec:background}). These expectations define the goal of explanations under the GDPR and thus describe requirements for explanatory content that should be \quotes{adapted to the expertise of the stakeholder concerned (e.g. layperson, regulator or researcher)} \cite{hleg2019ethics}. 
	Analysing these requirements we found a minimal set of explanation types that are necessary to meet these expectations: causal, justificatory and descriptive.
	
	%Does any solution exist in literature for providing adapted (user-centred) explanations aligned with the GDPR expectations?
	Most of the literature on AI and explanations (e.g. eXplainable AI) is currently focused on one-size-fits-all approaches usually able to produce only one of the required explanation types: causal explanations.
	%Probably this is the reason why we have not found any generic-enough tool for producing explanations of ADM under the GDPR.
	In this paper we take a strong stand against the idea that static, one-size-fits-all approaches to explanation (explainability) have a chance of satisfying GDPR expectations and requirements.
	In fact, we argue that one-size-fits-all approaches (in the most generic scenario) may suffer the curse of dimensionality. For example, a complex big-enough explainable software can be super hard to explain, even to an expert, and the optimal (or even sufficient) explanation might change from expert to expert. 
	This is why we argue that an explanatory tool for complex data and processes has to be user-centred and thus interactive.
	The fact that every different user might require a different explanation does not imply that there might be no unique and sound process for constructing user-centred explanations.
	In fact we argue that every interactive explanatory tool defines and eventually generates an Explanatory Space (ES) and that an explanation is always a path in the ES. 
	The explainee explores the ES, tracing a path in it, thus producing its own (user-centred) explanation.
	%This is why we are going to give a more formal definition of the ES and the other components of an explanatory process.
	
	This is why we assert that a more nuanced approach must be considered, where explanations are user-centred narratives that allow explainees to increase understanding through sense-making and articulation, in a manner that is fit to specified explainees, their goals, and their context of use.
	Upon these considerations, and following the High-Level Expert Group on Artificial Intelligence (AI-HLEG) Ethics Guidelines for Trustworthy AI, we propose here a new model of a User-Centred Explanatory Tool for Trustworthy AI, compliant with GDPR.
	To this extent we propose a definition of user-centred explanations as Explanatory Narratives, based on concepts drawn from ISO 9241. We present a formal model of an Interactive Explanatory Process consequently identifying 4 fundamental properties (the SAGE properties) of a good explanation and 3 heuristics for the exploration of the ES. Finally, we define the structure of the ES, combining the SAGE properties and the identified exploration heuristics, thus showing an application of the model to a real-case scenario.
	%We believe that, under these assumptions, the generation of tools allowing explainees to navigate through explanatory spaces can be made reasonably straightforward, and that evaluation dimensions such as objective \emph{efficacy} and subjective \emph{satisfaction} over the explanations generated by these tools can be assumed. 
	
	This paper is structured as follows: in Section \ref{sec:background} we provide some background information. %about explanation in literature, the GDPR, and the concept of transparency from the AI-HLEG Ethics Guidelines for Trustworthy AI. 
	In Section \ref{sec:narratives} we discuss over the GDPR, introducing our definition of Explanatory Narratives. In Section \ref{sec:related_work} we analyse existing work. While in Section \ref{sec:model} we propose a simple model of a User-Centred Narrative Explanatory Process. Finally in Section \ref{sec:discussion} we discuss the strengths and the weaknesses of our model, and in Section \ref{sec:conclusion} we conclude pointing to future work.
	
	\section{Background} \label{sec:background}
	%In this section, an informal but still reasonably precise summary of
	%\begin{inparadesc}
	%\item \quotes{explanation in literature},
	%\item \quotes{the right to explanation},
	%\item \quotes{the AI-HLEG guidelines for trustworthy AI}
	%\end{inparadesc}
	%will be given as far as needed for present purposes.
	
	\subsection{Explanations in Literature}
	In literature, many types of possible explanations have been thoroughly discussed, and it is not clear whether a complete and detailed taxonomy may exist. %\cite{lucas2006reason}
	In the field of Explainable Artificial Intelligence (XAI), the most discussed type of explanations is probably the causal one. 
	We can say that explanations can be causal or non-causal.
	Causal explanations may have many different shapes and flavours \cite{miller2018explanation}, including explanations based on \textit{causal attributions} (or chains), on \textit{causal reasoning}, etc..
	%According to \cite{hilton2007course}, at least 5 different types of causal chains exist:
	%\begin{inparadesc}
	%\item Temporal,
	%\item Coincidental,
	%\item Unfolding,
	%\item Opportunity chains,
	%\item Pre-emptive.
	%\end{inparadesc}
	%While, according to \cite{miller2018explanation}, at least 3 different types of explanations based on causal reasoning exist:
	%\begin{inparadesc}
	%\item Associative,
	%\item Interventionist,
	%\item Causal Counterfactual.
	%\end{inparadesc}
	Similarly, non-causal explanations can be of several different types, including (but not limited to): 
	\begin{itemize}
		\item Descriptive: explanations related to conceptual properties and characteristics: hypernyms, hyponyms, holonyms, meronyms, etc.
		\item Justificatory: explanations of why a decision is good.%, without details about the actual decision-making process \cite{biran2017explanation}.
		\item Deontic: justifications of the decision based on permissions, obligations and prohibitions \cite{meyer1993deontic}. In this sense, deontic explanations are a subset of justificatory explanations.
		%\item Teleological: explanations on the purpose of the decision by looking at its results.
		\item Contrastive: counterfactual explanations on events instead of the causes of events.
		%\item etc...
	\end{itemize}
	
	\subsection{The GDPR and the Right to Explanation} \label{sec:gdpr}
	%The General Data Protection Regulation (GDPR) is an important 2016 EU regulation on data protection and privacy. Since the GDPR is technology neutral, it does not directly reference Artificial Intelligence (AI), but several provisions are highly relevant to the use of AI for decision-making.
	The GDPR is technology-neutral, so it does not directly reference AI, but several provisions are highly relevant to the use of AI for decision-making \cite{ico2019}. 
	The GDPR defines the \quotes{Right to Explanation} as a right that individuals might exercise when their legal status is affected by a solely automated decision. In order to put the user in the conditions to be able to contest an automated decision and thus to exercise the right to explanation, the insights of the decisions have to be properly explained.
	
	The GDPR defines (indirectly) two modalities of explanation: explanations can be offered before (\emph{ex-ante}; artt. 13-14-15) or after decisions have been made (\emph{ex-post}; art. 22, paragraph 3). For each modality, the GDPR defines goals and purposes of explanations, thus providing a set of explanatory contents.
	From a technical point of view, there are technology-specific information to consider in order to fully meet the GDPR explanation requirements. 
	Fundamentally, ex-ante we should provide information that guarantees the transparency principle, such as describing:
	\begin{itemize}
		\item The algorithms and models pipeline composing the ADM.
		\item The data used for training (if any), developing and testing the ADM.
		\item The background information (e.g. the jurisdiction of the ADM).
		\item The possible consequences of the ADM on the specific data subject.
	\end{itemize}
	Ex-post the data subject should be able to fruitfully contest a decision, so he/she should be given access to: 
	\begin{itemize}
		\item The justification about the final decision.
		\item The run-time logic flow (causal chain) of the process determining the decision.
		\item The data used for inferring.
		\item Information (metadata) about the physical and virtual context in which the automated process happened.
	\end{itemize}

	%Furthermore, according to \cite{wachter2017counterfactual}  there is no clear link that suggests that explanations under Recital 71 (of the GDPR) require opening the black box but there is a clear indication \quotes{to obtain an explanation of the decision reached after such assessment and to challenge the decision}.
	In this scenario, law and ethics scholars have been more concerned with understanding the internal logic of decisions as a means to assess their lawfulness (e.g. prevent discriminatory outcomes), contest them, increase accountability generally, and clarify liability. 
	%\cite{wachter2017counterfactual} have assessed three purposes of explanations of automated decisions from the view of the data subject under the GDPR:
	%\begin{enumerate}
	%	\item to inform and help the subject \textit{understand} why a particular decision was reached,
	%	\item to provide grounds to \textit{contest} adverse decisions,
	%	\item to understand what could be \textit{changed} to receive a desired result in the future, based on the current decision-making model.
	%\end{enumerate}
	For example, \cite{wachter2017counterfactual} propose counterfactuals as a reasonable way to lawfully provide \textit{causal} explanations under the GDPR’s right to explanation. \cite{wachter2017counterfactual} takes strength from the Causal Inference theory, in which counterfactuals are hypothesised to be one of the main tools for Causal Reasoning \cite{pearl2019seven}.
	
	\subsection{Transparency and the AI-HLEG Guidelines for Trustworthy AI}
	The AI-HLEG has been charged by the European Union to identify a set of Ethics Guidelines for Trustworthy AI, published in April 2019 \cite{hleg2019ethics}.
	The AI-HLEG vision for a user-centred AI appears to incorporate the GDPR principles, trying to expand them into a broader framework based on 4 consolidated ethical principles, including:
	\begin{inparadesc}
		\item respect for human autonomy, 
		%\item prevention of harm, 
		\item fairness, 
		\item explicability.
	\end{inparadesc}
	From the aforementioned principles they derive seven key requirements for Trustworthy AI, including:
	\begin{inparadesc}
		%\item human agency and oversight, 
		%\item technical robustness and safety, 
		%\item privacy and data governance, 
		\item transparency, 
		\item diversity, non-discrimination and fairness, 
		%\item environmental and societal well-being 
		\item accountability.
	\end{inparadesc}
	
	The ethical principle of Explicability \cite{floridi2018ai4people} is associated to the requirements of Transparency and Accountability. The Transparency requirement, in turn, is clearly inspired by articles 13-14-15-22 of the GDPR.
	In a way, the AI-HLEG applies the technologically neutral GDPR by defining relevant guidelines on how Transparency can be achieved in Trustworthy AI systems, also through accessibility and universal design.
	The \quotes{Accessibility and Universal Design} requirement puts user-centrality at the core of Trustworthy AI systems. While the Transparency requirement encompasses transparency of elements relevant to an AI system (
	\begin{inparadesc}
		\item the data,
		\item the system,
		\item the business models 
	\end{inparadesc}
	), including: 
	%\begin{itemize}
	%	\item Traceability: \quotes{the data sets and the processes that yield the AI system’s decision, including those of data gathering and data labelling as well as the algorithms used, should be documented}.
	%	\item Explainability: is defined as \quotes{the ability to explain both the technical processes of an AI system and the related human decisions (e.g. application areas of a system)}, thus explicitly implying also \quotes{business model transparency}.
	%	\item Communication: \quotes{AI system’s capabilities and limitations should be communicated to AI practitioners or end-users in a manner appropriate to the use case at hand}.
	%\end{itemize} 
	\begin{inparadesc}
		\item Traceability,
		\item Explainability,
		\item and Communication
	\end{inparadesc} 
	
	\section{GDPR-compliant interactive explanatory tools for Trustworthy AI} \label{sec:narratives}	
	In this section we will discuss over the GDPR and the AI-HLEG guidelines, identifying a minimal set of required explanation types, thus proposing a new User-Centred Explanatory Tool based upon a definition of Explanatory Narratives aligned to ISO 9241.
	
	\subsection{Explanations under GDPR}
	The GDPR clearly draws a set of expectations to meet, in order to guarantee the Right to Explanation. These expectations are meant to define the goal of explanations and thus an explanatory content that may evolve together with technology. This explanatory content identifies at least 3 different types of explanations: 
	\begin{inparadesc}
		\item causal,
		\item descriptive,
		%\item teleological,
		\item justificatory.
	\end{inparadesc}
	We will refer to them as the minimal set of explanations required, for explaining ADM under the GDPR.%\footnote{Other types of explanations might be offered as well (e.g. contrastive explanations).}
	In fact, in the case of GDPR, we see that:
	\begin{itemize}
		\item Descriptive explanations are mostly required in the \emph{ex-ante} phase, to explain business-models, the possible effects of ADM on user, and characteristics and limitations of the algorithms.
		\item Causal explanations are mostly required in the \emph{ex-post} phase, to explain the causes of a solely automated decision.
		\item Justificatory explanations are required in both the \emph{ex-ante} and \emph{ex-post} phase, to justify decisions through permissions, obligations and so on.
	\end{itemize}
	The aforementioned explanations can be provided to the user through one or more explanatory tools as part of the whole AI system.
	This is why the AI-HLEG has defined some characteristics that these AI systems (and consequently their explanatory tools) should possess for trustworthiness. These characteristics include (among other things) transparency and user-centrality.
	
	\subsection{User-Centred Explanatory Tools}
	According to the AI-HLEG and the ICO \cite{ico2019}, user-centrality implies that (in the most generic scenario) explanations following a One-Size-Fits-All approach (OSFA explanations) are not user-centred by design. 
	For example, static symbolic representations where all aspects of a fairly long and complex computation are described and explained are one-size-fits-all explanations. 
	OSFA explanations have intuitively at least two problems.
	The first problem is that if they are small-enough to be simple, it is impossible that in a complex-enough domain they would contain enough information to satisfy the explanation appetite of every user. 
	The second problem is that if they contain all the necessary information, in a complex-enough domain they would contain an enormous amount of information and every user interested in a specific fragment of the explanation might look for it within hundreds of pages of explanations mostly irrelevant to her/his purposes.
	
	At this point one might observe that OSFA explanations could be useful for simple domains, but according to \cite{raymond2019culture} the complexity of a domain is exactly what motivates the need for explanations.
	In other terms, explanations are more useful to be given in complex domains.

	What are OSFA explanations?
	Static explanations are OSFA explanations by design, but sometimes OSFA explanations can also be interactive.
	In fact, intuitively, simply adding naive ways of interaction to a static explanatory tool does not imply that the new interactive tool is no more following a one-size-fits-all approach.
	This is why we argue that interactivity is not a sufficient property for user-centred explanatory tools.
	%We need more flexible tools than those based on a one-size-fits-all approach.
	
	What are the sufficient properties for user-centred explanatory tools?
	A user-centric explanatory tool requires to provide goal-oriented explanations. Goal-oriented explanations implies explaining facts that relevant to the user, according to its background knowledge, interests and other peculiarities that make her/him a unique entity with unique needs that may change over time.
	If the explanations have to be adapted to users, does this imply that we should have a different explanatory tool for every possible user?
	
	\subsection{Explanatory Narratives} \label{sec:explanatory_narratives}
	The fact that every different user might require a different explanation does not imply that there might be no unique and sound process for constructing user-centred explanations.
	In fact we argue that every interactive explanatory tool defines and eventually generates an Explanatory Space (ES) and that an explanation is always a path in the ES. 
	The explainee explores the ES, tracing a path in it, thus producing its own (user-centred) explanation.
	We are going to give a more formal definition of the ES and the other components of an explanatory process, later.
	
	Actually, being able to construct useful explanations is one of the main challenges of making science.
	This is why a lot of literature exist on how to construct scientific explanations.
	Constructing scientific explanations and participating in argumentative discourse are seen as essential practices of scientific inquiry (e.g., \cite{driver2000establishing}), that according to \cite{berland2009making} involves 3 different practices:
	\begin{inparadesc}
		\item sense-making,
		\item articulating,
		\item evaluating.
	\end{inparadesc}
	In fact a scientist should use evidence to make sense of phenomenon, articulating understandings into explanatory narratives. These explanatory narratives should be validated, e.g., defending them in a public debate against the attacks of scientific community.
	We believe that similarly to scientific explanations, constructing lawful explanations involves the same practices.
	For example, legal evidential reasoning can be seen as reasoning on evidences (sense-making) in order to justify/prove an hypothesis (articulating), in a way that the resulting arguments can be defended from opponents and accepted by judges during a debate (evaluating) \cite{prakken2019argumentation}.
	This is why we argue that a user-centred explanatory tool should be an instrument for sense-making, articulating and evaluating information into an explanatory narrative.
	If we focus on user-centred explanatory tools, then we are focusing on tools for sense-making through creating an explanatory narrative.
	What is an explanatory narrative?
	We consider an explanatory narrative as a sequence of information (explanans) to increase understanding over explainable data and processes (explanandum), for the satisfaction of a specified explainee that interacts with the explanandum having specified goals in a specified context of use.
	Our definition takes inspiration from \cite{passmore1962explanation,norris2005theoretical,lipton2001good}, integrating concepts of usability defined in ISO 9241, such as the insistence on the term \quotes{specific}, the triad \quotes{explainee}, \quotes{goal} and \quotes{context of use}, as much as the identification of a specific quality metric, which in our case are effectiveness and satisfaction. 
	The qualities of the explanation that provide the explainee with the necessary satisfaction, following the categories provided by \cite{norris2005theoretical}, can be summarized in a good choice of narrative appetite, structure and purpose. 
	
	The problems of a user-centred approach is that fully-automated explainers are unlikely to target quality parameters that guarantee the satisfaction of each specified explainee. Even if an AI could be used to generate such user-centred explanations, this would only shift the problem of explaining from the original ADM to another ADM -- the explanatory AI used to explain the original ADM. As such we believe that, in the case of user-centred explanations, the simplest solution is to require that reader (explainee) and narrator (explainer) are the same, generating the narration for themselves by selecting and organizing narratives of individual event-tokens according to the structure that best caters their appetite and purpose. 
	In this sense, a tool for creating explanatory narratives would allow users to build intelligible sequences of information, containing arguments that support or attack the claims underlying the goal of a narrative explanatory process defined by explainee/explainer and explanandum.
	
	\section{Related work} \label{sec:related_work}
	Apparently most of the tools for explaining ADM are static (e.g. AIX360 \cite{arya2019one}).
	Interactive tools also exist \cite{fox2017explainable,cocarascu2019extracting,vcyras2019explanations,zhong2019explainable}, but:
	\begin{inparaenum}[1)]
		\item they do not consider to offer descriptive explanations, but other types of explanations on pre-defined aspects of the ADM;
		\item or they generate explanations automatically (e.g from static argumentation frameworks), using templates.
	\end{inparaenum}
	Completely automated sense-making or understanding articulation is possible only with very specific ADMs, or by pre-defining narrative scenarios that can be as powerful as dangerous \cite{bennet1981reconstructing,verheij2015arguments}.
	%, thus relying on important assumptions on the content of the explanations and thus on the expertise of the stakeholder concerned.
	Furthermore, a solely automated explainer is an ADM process itself, that might require to be explained as well.
	
	Actually, as far as we know, there is no tool for explanatory narratives of ADMs, but more generic tools for teaching exist such as \cite{suthers1997integrated,sandoval2004explanation}.
	There are some interesting similarities between our work and the two aforementioned works, including
	\begin{inparadesc}
		\item the assumption that highlighting the structural elements of the explanandum is necessary in the articulation of an understanding.%, such as evidence , will help to recognize the importance of various types of knowledge for sense-making.
		%\item commands that can be used to communicate user-centred explanations in a way that is consistent with the norms of the scientific community.
	\end{inparadesc}
	
	\section{A User-Centred Narrative Explanatory Process} \label{sec:model}	
	In this section we present  our model of User-Centred Narrative Explanatory Process under the GDPR. We start modelling a generic explanatory process, giving a formal definition of  explanandum, explanans and Explanatory Space. Concurrently to modelling we show, step-by-step, an application of the model in a real-case scenario.
	
	\subsection{Real-Case Scenario}
	We present here a real-case scenario we will continuously refer to when defining our user-centred narrative explanatory process.
	This real-case is about the conditions applicable to child's consent in relation to information society services. The art. 8 of GDPR fixes at 16 years old the maximum age for giving the consent without the parent-holder authorization. This limit could be derogated by the domestic law. In Italy the legislative decree 101/2018 defines this limit at 14 years. In this situation we could model legal rules in LegalRuleML \cite{athan2013oasis,palmirani2018modelling} using defeasible logic, in order to be able to represent the fact that the GDPR art. 8 rule is overridden with the Italian's one. 
	The SPINDle legal reasoner processes the correct rule according to the jurisdiction (e.g., Italy) and the age. 	
	Suppose that Marco (a 14 years old Italian teenager living in Italy) uses Whatsapp, and his father, Giulio, wants to remove Marco's subscription to Whatsapp because he is worried about the privacy of Marco when online. 
	In this simple scenario, the ADM system would reject Giulio's request to remove Marco's profile.%, because of the Italian legislative decree 101/2018.
	What if Giulio wants to get an explanation of the automated decision?
	To answer this question we have to pick an explanatory process. 
	
	\subsection{The Interactive Explanatory Process}
	%\begin{wrapfigure}{R}{.5\columnwidth}
	\begin{figure}[h]
		\vspace{-10pt}
		\centering
		\includegraphics[width=.75\columnwidth]{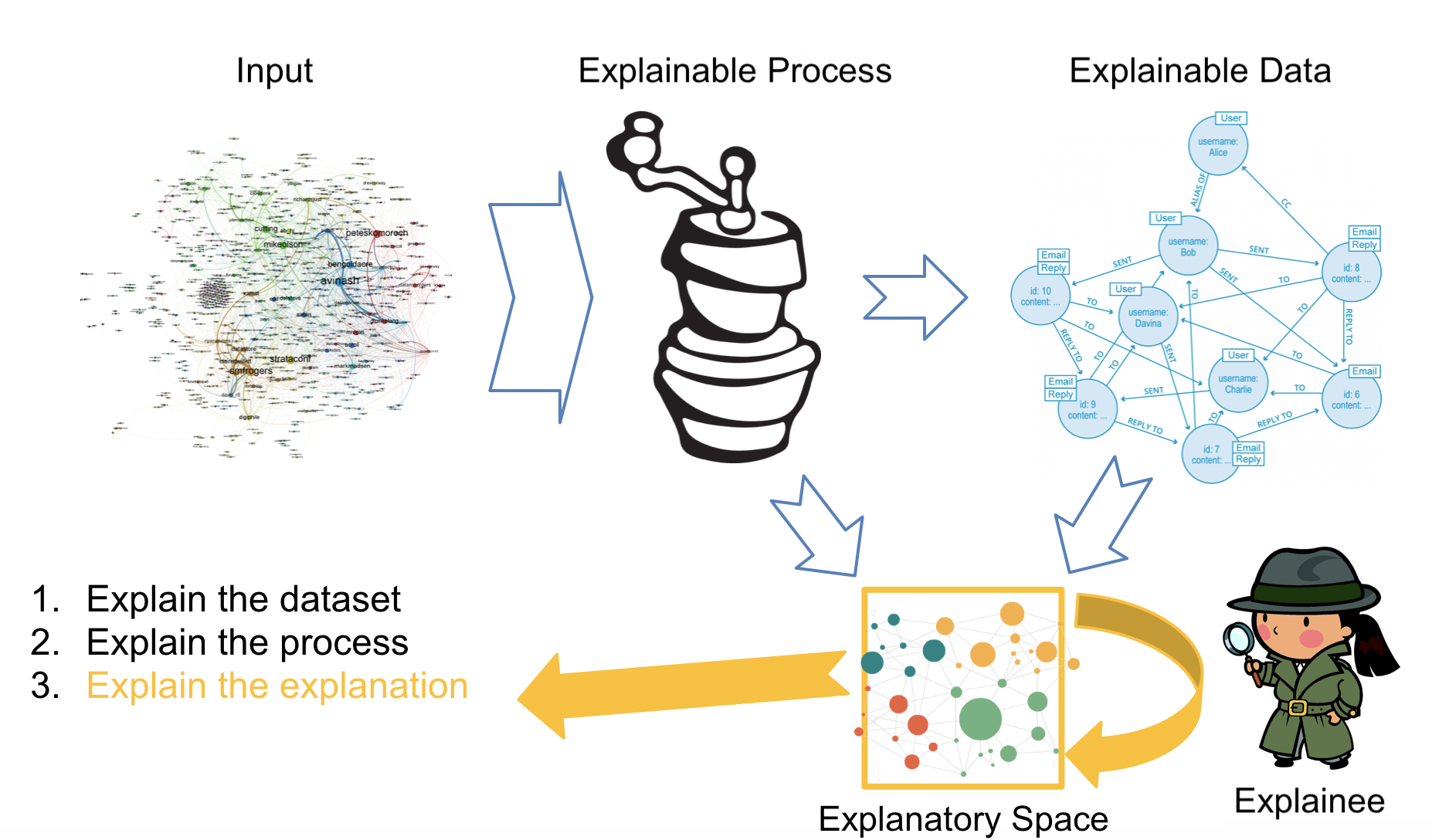}
		\caption{Stylized interactive explanatory process.}
		\label{fig:explanatory_process}
		\vspace{-10pt}
	%\end{wrapfigure}
	\end{figure}
	According to the definition given in Section \ref{sec:explanatory_narratives}, a user-centred (interactive) narrative explanatory process is a process explaining an explanandum to an explainee (reader and narrator), thus producing as output an explanans (explanatory narrative) that is meaningful for the specific explainee.
	As shown in figure \ref{fig:explanatory_process}, an explanatory process is a function $p$ for which $p\left(D, E_t, i_t\right) = E_{t+1}$, where:
	\begin{itemize}
		\item $D$ is the explanandum.
		\item $E_t$ is the input explanans, at step $t \geq 0$.
		\item $i_t$ is the user interaction at step $t$.
		\item $E_{t+1}$ is the output explanans, thus at step $t+1$.
	\end{itemize}
	We can iteratively apply $p$, starting from an initial explanans $E_0$, until satisfaction. 
	The user interaction $i$ is a tuple made of an action $a$ taken from the set $A$ of possible actions, and a set of auxiliary inputs required by the action $a$.
	
	\subsection{The Explanandum}
	As defined by AI-HLEG guidelines, in this setting the explanandum is a collection of context-dependent information, made of one or more:
	\begin{inparadesc}
		\item datasets,
		\item processes,
		\item business models (that are higher-order processes).
	\end{inparadesc}
	Naturally, the explanandum has to be \emph{explainable} in order to be explained.
	In this scenario, datasets and processes are said to be explainable when they have a clear and not ambiguous symbolic representation, i.e., when their data items and rules are aligned with meaningful ontologies\footnote{These ontologies do not have necessarily to be explicit, formal or complete.} for the end-users.
	The context-dependency implies that, in the most generic scenario, the information contained in datasets and processes is not sufficient to understand their nature without external knowledge. This external knowledge is commonly assumed to be part of the explainee knowledge (e.g. the knowledge about how to interpret a natural language).
	We will refer to this external common knowledge as the \emph{explanandum context}, considering it as a dataset implicitly part of the explanandum. %This implicit and large dataset has to be somehow aligned to the explicitly given datasets and processes. We assume that this is possible through word sense disambiguation. 
	
	In our real-case the explanandum is made of:
	\begin{inparadesc}
		\item a rule-base (the LegalRuleML one),
		\item a dataset of premises (the information about the involved entities),
		\item a dataset of conclusions obtained by applying the premises to the rule-base,
		\item a causal chain (the ordered chain of rules involved in the production of the dataset of conclusions).
	\end{inparadesc}
	
	What are datasets and processes? \\
	A dataset is a tuple $\left<X, N, S\right>$ where:
	\begin{itemize}
		%\item $K$ is the original representation of data-items.
		\item $X$ is a set of data-items.
		\item $N$ is the set of (possibly informal) ontologies describing $X$.
		%\item $Z$ is the set of ontologies (if any) different from $N$ but implicitly related to $N$.
		\item $S$ are the (possibly unstructured) sources used to derive $N$ and $X$.
	\end{itemize}
	A process is a tuple $\left<D_1, D_2, D_3\right>$ where:
	\begin{itemize}
		\item $D_1$ is the dataset of the process \textit{inputs}, i.e. the domain.
		\item $D_2$ is the dataset of the process \textit{function}.
		\item $D_3$ is the dataset of the process \textit{outputs}, i.e. the codomain.
	\end{itemize}
	When $D_1$ is a collection of processes, the process is said to be higher-order.
	
	In our real-case the sources of the function/rule-base are the GDPR (art. 8) and the Italian legislative decree 101/2018, while the source of the inputs/premises and the outputs/conclusions is the textual description of the case.
	The main ontology of the rule-base is a representation of the knowledge behind the SPINDle legal reasoner, while the ontologies for the premises and the conclusion might include the Dublin Core Schema, etc..
	
	\subsection{The Explanans and The Explanatory Space}
	The explanans is a particular type of dataset, in fact it is an ordered sequence of arguments (contained in the explanandum) useful to the explainee to reach its goals.
	The set of all the possible explanans reachable by an explainee interacting with the explanatory process, given an explanandum and an initial explanans, is called \emph{Explanatory Space} (ES).
	The explanatory space is defined by: 
	\begin{itemize}
		\item $p$ (the explanatory process),
		\item $A$ (the set of actions),
		\item $D$ (the explanandum), 
		\item the initial explanans $E_0$.
	\end{itemize}
	%The explanatory space is independent from the choices taken by the explainee during the explanatory process, unless the explainee is able to modify the explanandum.
	%\subsubsection{Properties of a good Explanation}
	Thus, following the formalizations previously presented, the components of an explanatory process are the:
	\begin{inparadesc}
		\item Explainee,
		\item Instances ($X$),
		\item Ontologies ($N$),
		\item Sources ($S$).
	\end{inparadesc}

	This is why we say that a good explanans (explanation) has to be bound to both the explainee and the explanandum. In other words, a good ES is:
	\begin{itemize}
		\item Sourced: bound to the sources.
		\item Adaptable: bound to the specific explainee.
		\item Grounded: bound to the instances.
		\item Expandable: bound to the ontologies.
	\end{itemize}
	We will call to these properties: the SAGE properties of a good ES.
	
	\subsection{Heuristics for Exploring the Explanatory Space}
	Assuming that the explanandum $D$ is provided as defined above, then:
	\begin{enumerate}
		\item How do we define the explanatory process $p$?
		\item How do we pick the initial explanans $E_0$?
		\item How do we pick the set of actions $A$?
	\end{enumerate}
	The answers to these questions are highly dependent to the constraints that the explanatory space is supposed to have.
	In our scenario, we want to define an explanatory space sufficient to provide user-centred explanations of ADMs for trustworthy AI.
	
	Considering that the narrative explanatory process $p$ consists in exploring an Explanatory Space, in order to define $p$ we identify 3 policies for exploring the ES, namely: 
	\begin{inparadesc}
		\item the \textit{Simplicity},
		\item the \textit{Abstraction},
		\item and the \textit{Relevance} policies.
	\end{inparadesc}
	%\fixme{Missing definition of relevance/novelty.}
	
	\textbf{Simplicity} is mentioned in recommendation 29.5 of the AI-HLEG Policy and Investment Recommendations \cite{hleg2019policy}: \quotes{Ensure that the use of AI systems that entail interaction with end users is by default accompanied by procedures to support users [...]. These procedures should be accompanied by simple explanations and a user-friendly procedure}.
	In fact, the amount of information that can be effectively provided to a human is limited by physical constraints. Thus simple explanations about something are more likely to be accepted and understood and they are better than complicated ones and should be presented earlier than complex ones.
	%\fixme{Missing a precise definition of simplicity.}
	%In this scenario we might expect that simple explanations are good for non-expert users, while complex explanations may be good for expert users.
	%A complicated or even complex explanation can be overwhelming and harmful for a non-expert user.
	%This is why we believe that a good heuristic for exploring the explanatory space would present simple explanations before complex ones.
	%We will refer to this heuristic as the simplicity policy.
	
	%Following the same intuition that has driven us to the simplicity policy, we define the abstraction policy.
	The \textbf{Abstraction} policy maintains that preferring abstract concepts over their concretizations helps in keeping the explanation as direct as possible. This is because there might be many concretizations of the same abstract concept. In other words, exploring concretizations before abstractions would generate longer paths in the explanatory space rather than the opposite. If we explore abstractions before concretizations, it is more likely that the ES is less complicated, without losing informative content.
	
	The \textbf{Relevance} policy bounds further the explanatory process to the purposes/objectives of the explainee. It states that the information that is more likely to be relevant for the explainee (to reach its objective) should be presented earlier than the less relevant one. One of the expected effects of the relevance policy is that explanations will be shorter.
	
	The Simplicity, Abstraction and Relevance policies shape the explanatory process $p$, putting significant constraints on the initial explanans $E_0$ and the set of actions $A$.
	Simplicity implies that in the explanatory process we start from a very minimal explanans and iteratively we add information to it in order to make it more detailed and complex. Simplicity also implies that simple representations of the explanandum (e.g. natural language descriptions) should be presented before the original representations (the \quotes{Ground}; e.g. XML, JSON, etc..).
	Abstraction implies that the explanans should start from generic and conceptual information about the explanandum (e.g. its size and other meta-data), going toward more specific and concrete information (the \quotes{Ground}).
	Relevance implies (among other things) that $E_0$ should contain the most relevant information possible, and that further details should be firstly about the entities directly involved in $E_0$. The other entities should be explored/presented in a second moment, if needed.
	
	\subsection{The Initial Explanans}
	The initial explanans $E_0$ should contain an overview of the underlying Explanatory Process (EP), giving explicit information about the purposes (e.g. to give insights about a sequence of events, to verify an hypothesis, etc..) of the EP.\footnote{It is not excluded that the original purposes might change during the explanatory process.}
	$E_0$ should also provide an overview of the explanandum, pointing to information about the metadata (e.g. size, language, knowledge representation conventions). 
	We will refer to the information contained in $E_0$ as the \quotes{incipit} or \quotes{background} of the EP.
	
	%The initial explanans contains information about the context and the goals of the explanatory process.
	In the case of explanations under the GDPR, $E_0$ should specify (among other things) whether the EP is operating ex-ante or ex-post, and the explanandum referred by $E_0$ should contain all the information mentioned in Section \ref{sec:gdpr}.  
	A succinct justification about the automated decision (required in the ex-post phase) can be generated through a static explanatory tool such as AIX360 \cite{arya2019one}, and should be given as part of the EP overview. 
	
	In our real-case, the justification about the automated decision states that Giulio's request has been rejected because of the Italian decree.
	
	\subsection{The High-Level Actions and the Structure of the Explanatory Space}
	%\begin{wrapfigure}{R}{.5\columnwidth}
	\begin{figure}
		\vspace{-10pt}
		\centering
		\includegraphics[width=.75\columnwidth]{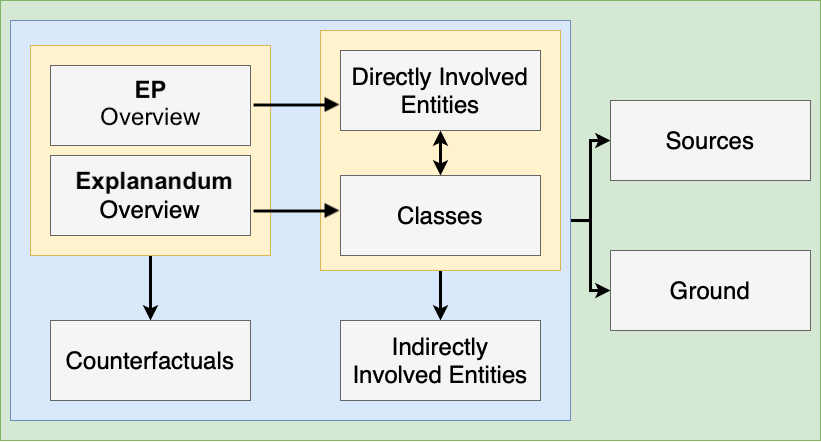}
		\caption{Structure of the Explanatory Space: arrows show the flow of information, while every rectangle represents a different sub-stage in the Explanatory Space.}
		\label{fig:explanatory_space_structure}
		\vspace{-10pt}
	%\end{wrapfigure}
	\end{figure}
	The explainee explores the Explanatory Space through a set of pre-defined actions meant to be used to build explanations respecting the SAGE properties.
	%The SAGE properties can be regulated by the explainee through a set of distinct commands/actions at each stage of the explanation structure.
	For every SAGE property, we identify a set of commands that may be used by the explainee during the Explanatory Process:
	\begin{itemize}
		\item \textbf{Source-ability} (commands): used to show the source of a model (a law, a paper, etc..).
		\item \textbf{Adapt-ability}: used to keep track of important information while exploring the explanatory space, building an argumentation framework.
		\item \textbf{Ground-ability}: used to refer and show specific parts of the explanandum in their original format (e.g. XML, JSON, SQL, etc..), and used for generating counterfactuals.
		\item \textbf{Expand-ability}: used to deepen concepts, aligning them to other concepts available in the explanandum context. 
	\end{itemize}
	
	Defining the explanatory process $p$, the (high-level) set of commands/actions $A$ and the initial explanans $E_0$, we have also defined the structure of the Explanatory Space.
	%The structure of the explanatory space may impact drastically on the user-experience. If information would be structured in a proper manner, the user might be able to easily explore it, looking for a tailored explanation. On the other side, if information would be badly structured, the user might be overwhelmed by useless information without being able to find any useful explanation.
	The resulting structure (shown in figure \ref{fig:explanatory_space_structure}) is made of 6 main stages:
	\begin{inparadesc}
		\item Incipit (EP and Explanandum Overview),
		\item Core Information (Classes and Directly Involved Entities),
		\item Marginal Information (Indirectly Involved Entities),
		\item Ground,
		\item Sources,
		\item Counterfactuals.
	\end{inparadesc}	
	Each stage of the structure is involved in a different step of the explanans construction.
	
	In our real-case the explainee (Giulio) wants to get an explanation of SPINdle's conclusions (the process' decision).
	Giulio uses the Explanatory Process (EP) on the explanandum. The EP starts from the \quotes{Incipit} stage (the initial explanans) showing:
	\begin{itemize}
		\item A succinct \emph{ex-post} justification of the process' decision, and that the justification is related to known concepts that can be easily explored.
		\item The explanandum is made of a knowledge base and a process having a set of inputs that can be changed by Giulio in order to understand whether the justification is valid and useful for him (maybe it is not).
	\end{itemize}
	The first level result of EP is too shallow, thus Giulio asks to EP to go down to the \quotes{Core Information} stage showing more information about the explanandum. 
	Now Giulio sees that the data is composed by: a set of logical conclusions, the hierarchy of rules used to get those conclusions (the causal chain), the premises on which the rules have been applied.
	Giulio wants to get more information about the rules, thus it moves to the \quotes{Marginal Information} stage, finding out that the Italian decree 101/2018 has rebutted the GDPR and it is responsible for the final decision, thus it marks that information as an argument that supports the process' decision.
	Furthermore, Giulio sees that every rule is ground-able to a LegalRuleML component, and linkable to the pertinent source of law that justifies the rule.
	Giulio can also see rebuttals, and if he would ask the EP to tell more about the GDPR rebuttal he would find out that the \quotes{Lex specialis derogat generali} is applied, causing the activation of the rule associated to the Italian decree instead of the rule associated to the GDPR.
	
	\section{Discussions} \label{sec:discussion}
	With an explanatory tool based on our model, the user can explore the explanatory space and build its own explanatory narrative through a set of pre-defined actions. The explanatory narrative can be built through the adapt-ability commands, by defining an argumentation framework. The exploration of the explanatory space can be performed through the expand-ability, source-ability and ground-ability commands.
	The resulting tool is user-centred by design and can be used for finding evidence to make sense of phenomena (sense-making), articulating understandings into an explanatory narrative.
	Furthermore, we claim that the structure of Explanatory Space we identified is sufficient to produce the descriptive, causal and justificatory explanations required by the GDPR.
	In fact, assuming that the information at the \quotes{Background} stage defines the initial explanans, we have that:
	\begin{inparadesc}
		\item Descriptive explanations can be obtained by reasoning over the \quotes{Core and Marginal Information} stages of the ES,
		\item Causal explanations over the \quotes{Counterfactuals} stage mainly,
		\item and Justificatory explanations over the \quotes{Incipit} and \quotes{Sources} stages mainly.
	\end{inparadesc}
	It is possible to apply logical rules in order to automatize the production of reasonable explanations.
	In fact, Description Logic can be used for reasoning over descriptions, 
	Causal Inference \cite{pearl2019seven} for reasoning over causations, Defeasible Deontic Logic for reasoning over justifications.
	%keeping in mind that automated explanatory systems are ADMs that have to be explained as well.
	
	Despite this, our model has some limitations.
	Because it is meant to be effective in building user-centred explanations, but not community-centred ones. In fact, constructing strong, effective and non-over-fitted explanations is historically an iterative and community-centred process. 
	
	\section{Conclusions and Future Work} \label{sec:conclusion}
	%\fixme{What does your theoretical formulation contribute to?}
	We have introduced a model of a User-Centred Narrative Explanatory Process, based on concepts drawn from ISO 9241, as a promising contribution to Trustworthy AI compliant with the GDPR.
	To this end, we identified a minimal set of required explanation types and we converted the problem of generating explanations into the identification of an appropriate path over an Explanatory Space (ES) defined and eventually generated by every user-centred (interactive) explanatory tool.
	Finally we provided a reasonable structure of the ES through the identification of the SAGE properties and of some space exploration heuristics.
	%We can see our tool as a tool for manually building argumentations graphs for trustworthy AI and constructing valid explanations, under the GDPR, through sense-making and articulation of explanatory narratives.
	Building a working prototype based upon our model is the next natural step. We are also considering to extend the current model in order to make it suitable also for community-centred explanatory tools.
	
	\bibliographystyle{splncs04}
	\bibliography{document}
	
\end{document}